\documentclass[aps,prl,twocolumn,showpacs,groupedaddress,affilletter,nobibnotes,nofootinbib]{revtex4}

\usepackage{graphicx}
\usepackage{epsfig}

\newcommand{\columnwidthstrach}{3.5in}

\begin{document}


\title{Two Distinct Time-Scale Regimes of the Effective Temperature for an Aging Colloidal Glass}


\author{D.~R. Strachan}
\thanks{Email: drstrach@sas.upenn.edu
\\ Current address: Deptartment of Physics and Astronomy, University of
Pennsylvania, Philadelphia, PA 19104}
\affiliation{Center for
Superconductivity Research, University of Maryland, College Park,
MD 20742}
\author{G. C. Kalur}
\thanks{Current address: IRIX Pharmaceuticals, Inc, Florence, SC - 29501}
\author{S. R. Raghavan}
\affiliation{Department of Chemical \& Biomolecular Engineering,
University of Maryland, College Park, MD 20742}



\date{\today}

\begin{abstract}
Colloidal dispersions of Laponite platelets are known to age
slowly from viscous sols to colloidal glasses. We follow this
aging process by monitoring the diffusion of probe particles
embedded in the sample via dynamic light scattering. Our results
show that the time-dependent diffusion of the probe particles
scales with their size. This implies that the
fluctuation-dissipation theorem can be generalized for this
out-of-equilibrium system by replacing the bath temperature with
an effective temperature. Simultaneous dynamic rheological
measurements reveal that this effective temperature increases as a
function of aging time and frequency. This suggests the existence
of two regimes: at probed time scales longer than the
characteristic relaxation time of the Laponite dispersion, the
system thermalizes with the bath, whereas at shorter time scales,
the system is out-of-equilibrium with an effective temperature
greater than the bath temperature.
\end{abstract}


\maketitle

For a system in thermodynamic equilibrium with a heat bath at
temperature $T_{bath}$, the well-known fluctuation-dissipation
theorem (FDT) relates the response under an external disturbance
to the random fluctuations that exist in the absence of the
disturbance~\cite{Kubo91_1}. Recently, there has been extensive
interest in extending this formulism to out-of-equilibrium systems
by incorporating an effective temperature $T_{eff}$  that differs
from $T_{bath}$~\cite{Cugliandolo93_1prl,Cugliandolo97_1pre}. The
allure of extending the FDT is that it could help to generalize
the behavior of some widely encountered out-of-equilibrium
materials, such as glasses, gels, and granular media, which evolve
or relax over very long time scales. Attempts to extend the FDT to
glassy
systems~\cite{Grigera99_1prl,Herisson02_1prl,Buisson03_1el},
granular materials~\cite{Anna03_1nat,Ojha04_1nat}, and to a
variety of soft
materials~\cite{Bellon01_1el,Bellon02_1pd,Bonn03_1jcp,Abou04_1prl}
have been reported. In particular, for an aging colloidal glass,
$T_{eff}$ has recently been found to increase with age for short
aging times~\cite{Abou04_1prl}. To test the validity of this
result, the dependence of $T_{eff}$ on the observation time scale
(inverse of the probed frequency $\omega$) needs to be
examined~\cite{Abou04_1prl,Cipelletti05_1jpcm}. It is also
imperative to demonstrate that the generalized FDT has the same
form as the classic version, but with $T_{eff}$ replacing
$T_{bath}$.

In this Letter we demonstrate a generalized FDT for an aging
colloidal glass formed by nanoscale platelets of Laponite RD in
water. As the sample ages, we measure the diffusion of immersed
probe particles of different sizes via dynamic light scattering
(DLS). Significantly, we find the time-dependent diffusion of the
probe particles to scale with their size. This, in turn, implies
that the FDT can be generalized for this aging system by replacing
the bath temperature with a $T_{eff}$. We then determine the
dependence of $T_{eff}$ on aging time and frequency by combining
DLS measurements with simultaneously taken dynamic rheological
measurements. The results show that $T_{eff}$ increases as a
function of aging time and probe frequency. This suggests that
there are two distinct regimes of the effective temperature
relevant for this colloidal glass, corresponding to probed time
scales greater than and less than the characteristic relaxation
time of the system.

\begin{figure}
\epsfig{file=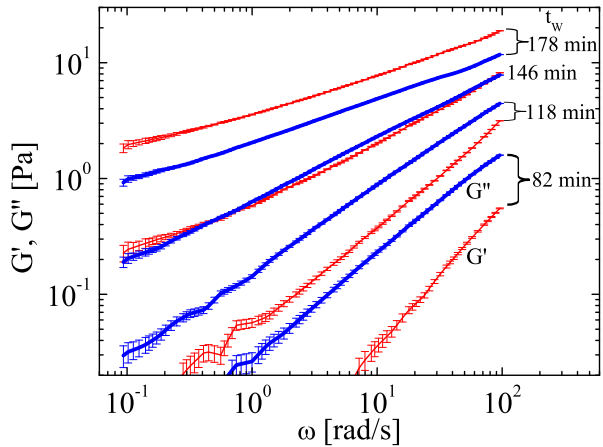,clip=,width=\columnwidthstrach}
\caption{\label{visco-elastic} Aging behavior of $G'$ (thin lines)
and $G''$ (thick lines) of a 3.0 wt \% Laponite suspension.  Data
are determined at set aging times by making measurements of the
viscoelastic modulus at 25 frequencies between 0.1 rad/s and 100
rad/s every 10 minutes. These data are then interpolated to
specific aging times and frequencies.}
\end{figure}

The system studied here consists of a colloidal dispersion of
Laponite RD platelets (roughly 3 nm thick and 25 nm in diameter)
immersed in an aqueous solution at pH 10. To start the experiment,
3 wt \% of Laponite RD is added to about 65 mL of water and
stirred for 20 min. The dispersion is then pushed through a 0.45
mm filter to break up any remaining particle aggregates. At this
stage, we mark time equal to zero for the Laponite aging
experiment~\cite{Bonn99_1lang,Knaebel00_1el,Abou01_1prb}. The
sample is then distributed among several light scattering cuvettes
and the couette of a Rheometrics strain-controlled (RDAIII)
rheometer for simultaneous studies via DLS and rheology. The
dynamic rheological response at different aging times is measured
at small strains well within the linear viscoelastic range so as
to ensure that the measurements do not disturb the aging. Figure\
\ref{visco-elastic} shows the evolution of the storage and loss
moduli, $G'(\omega)$ and $G''(\omega)$, with aging time $t_{w}$
for the Laponite dispersion. The sample is seen to evolve from a
viscous sol at short $t_{w}$ ($\sim$ 82 min) to an elastic medium
at large $t_{w}$ ($\sim$ 178 min). To simultaneously probe the
fluctuations of the system as it ages, small amounts of
polystyrene spheres (of diameter 50, 100, or 200
nm)~\cite{nano_spherres} are added to the light scattering
cuvettes. The intensity autocorrelation function is measured on
these samples with a Photocor-FC light scattering instrument
employing a 5 mW laser light source at 633 nm and a multi-tau
logarithmic correlator. Because simultaneous measurements are
performed on portions divided out from a single prepared
dispersion, we are able to ensure that the rheology and DLS
measurements correspond to the same aging time $t_{w}$.

\begin{figure}
\epsfig{file=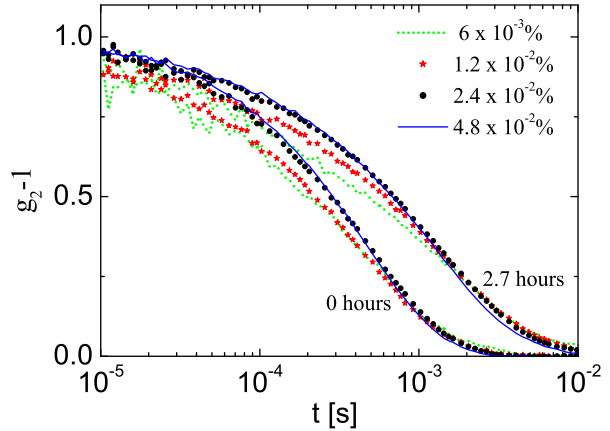,clip=,width=\columnwidthstrach}
\caption{The two sets of correlation functions, $g_2(\mathbf{q},t)
- 1$, are measured at two different aging times for four different
concentrations of 50 nm polystyrene spheres in a 3 wt \%
dispersion of Laponite. \label{concentration_vary} }
\end{figure}

Our intent with the DLS experiments is to measure the mean-squared
displacement of probe particles embedded in the Laponite
dispersion while the sample is aging. For this, we exploit the
fact that the scattering from the probes dominates over the
scattering from the Laponite particles themselves. This is shown
by Fig.\ \ref{concentration_vary}, where we plot the intensity
autocorrelation function $g_2$ taken at $90^{\circ}$ for different
concentrations of 50 nm probes. The correlation function is found
to be independent of probe concentration for concentrations of
0.02\% or higher. In this regime, we can assume the scattering to
be predominantly due to the probes, and thereby determine the
average mean-squared displacement of the probes through.
\begin{equation}
g_2(\mathbf{q},t) - 1= {{\langle i(\mathbf{q},0)i(\mathbf{q},t)
\rangle} \over {\langle i(\mathbf{q},0) \rangle}^2} - 1 =
f(A)|F_s(\mathbf{q},t)|^2 \label{correlation_function_at_detector}
\end{equation}
Here, the magnitude of the scattering wave vector $\mathbf{q}$ is
$q = {{4 \pi n} \over {\lambda}}sin(\theta / 2)$,
$F_s(\mathbf{q},t) = e^{(-q^2\langle \Delta r^2(t) \rangle/6)}$ is
the self-intermediate scattering function, $i(\mathbf{q},t)$ is
the intensity at the detector, and $f(A)$ is the spatial coherence
factor~\cite{Berne00_1}. At higher concentrations than shown in
Fig. 2, the addition of probe particles alters the aging of the
Laponite dispersion. This indicates that only a range of probe
particle concentrations will be large enough to dominate
scattering while not affecting the Laponite aging (e.g., for 50 nm
spheres, this corresponds to concentrations of roughly 0.02 to
0.06 wt \%~\cite{concentrations}).

We now use the determined $\langle \Delta r^2(t) \rangle$ of the
embedded probe particles to determine whether a generalization of
the FDT is compatible with the data for this out-of-equilibrium
system. By assuming that the bulk stress relaxation is the same as
the local stress relaxation affecting the probe particle, it has
been shown~\cite{Mason95_1prl} that probe particles immersed in a
viscoelastic medium can obey the generalized Stokes-Einstein
relation
\begin{equation}
a\langle \Delta\tilde{r}^2(s) \rangle = T \left({{k_B} \over {s
\tilde{G}(s) \pi}} \right).
\label{classic_stokes_einstein_relation}
\end{equation}
In the above, $s$ is the Laplace frequency, $\langle
\Delta\tilde{r}^2(s) \rangle$ is the Laplace transform of the mean
squared displacement of probe particle of radius $a$, and
$\tilde{G}(s)$ is the viscoelastic modulus. Equation\
(\ref{classic_stokes_einstein_relation}) can be generalized to
out-of-equilibrium systems by replacing the temperature $T$ with
an effective temperature $\Theta(s)$, yielding
\begin{equation}
a\langle \Delta\tilde{r}^2(s) \rangle = \Theta(s) \left({{k_B}
\over {s \tilde{G}(s) \pi}} \right).
\label{general_stokes_einstein_relation}
\end{equation}
This can be cast into a frequency-dependent form by substituting
$s=i \omega$, with the effective temperature given
by~\cite{Pottier03_1pa,Pottier04_1pa,Pottier05_1condmat}
\begin{equation}
T_{eff}(\omega) \mathrm{Im} \left[ {1 \over G^{*}(\omega)} \right]
= \mathrm{Im} \left[ {\Theta(\omega) \over G^{*}(\omega)} \right].
 \label{T_eff}
\end{equation}

The effective temperature $\Theta(s)$ of Eq.\
(\ref{general_stokes_einstein_relation}) accounts for deviations
from the equilibrium relation of Eq.\
(\ref{classic_stokes_einstein_relation}). If $\Theta(s)$ is to act
like a temperature it must be an intensive variable and
independent of the size $a$ of the probe particle. Assuming
$\tilde{G}(s)$ is also intensive, this would require that the term
$a\langle \Delta\tilde{r}^2(s) \rangle$ be independent of $a$. To
demonstrate the $a$ independence, we multiply Eq.\
(\ref{general_stokes_einstein_relation}) by $s/6$, which can be
arranged to yield
\begin{equation}
a\tilde{D}(s) = \Theta(s) \left({{k_B} \over {\tilde{G}(s) 6 \pi}}
\right). \label{general_stokes_einstein_relation_aging_test2}
\end{equation}
In the above we have used $\tilde{D}(s) = {{s \langle
\Delta\tilde{r}^2(s) \rangle} \over {6}}$, which is the Laplace
transform of the time-dependent diffusivity, $D(t) = {1 \over 6}
\cdot {{\partial \langle \Delta r^2(t) \rangle} \over {\partial
t}}$~\cite{Mason97_1josaa}. The inverse Laplace transform of Eq.\
(\ref{general_stokes_einstein_relation_aging_test2}) yields
\begin{equation}
aD(t) = f_w(t),
\label{general_stokes_einstein_relation_aging_test4}
\end{equation}
where $f_w(t)$ is an unspecified function of time that should be
independent of $a$ if the effective temperature acts as an
intensive variable.

\begin{figure}
\epsfig{file=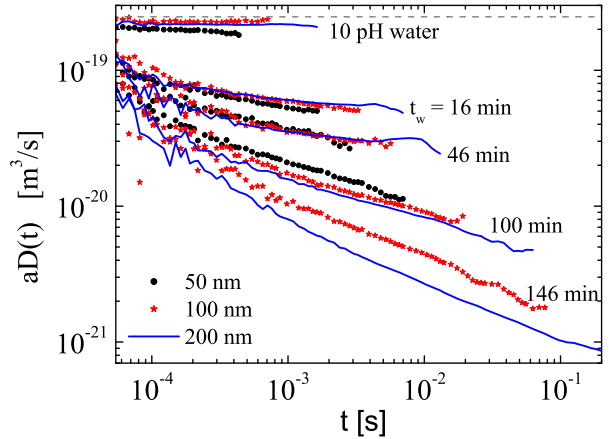,clip=,width=\columnwidthstrach}
\caption{Time dependent diffusion, $D(t)$, normalized with the
radius, $a$, of the probe particles.  The measurements were
performed in portions taken from the the same 3 wt \% Laponite
dispersion as used in the rheological measurements of Fig.\
\ref{visco-elastic}. The aging times are noted next to the curves.
A comparison to probe particles immersed in pH 10 water is shown
at the top of the plot.  The gray line is the value expected for
spheres immersed in water at 298 K. \label{scale_Dt} }
\end{figure}

To test Eq.\ \ref{general_stokes_einstein_relation_aging_test4},
we first determine $D(t)$ by numerically differentiating the
logarithm of the autocorrelation function, which eliminates the
coherence factor $f(A)$ in the analysis. As a check that this
yields a reasonable result, we analyzed the probe particles (at
the same concentrations) in pH 10 water without Laponite, which
gave the three marked lines in Fig.\ \ref{scale_Dt} and are in
reasonable agreement with the diffusion of spheres in water at 298
K, represented as a horizontal gray line. When attempted on the 3
wt \% Laponite solution from Fig.\ \ref{visco-elastic}, $aD(t)$
again shows good agreement between the various probe sizes at
early $t_{w}$, but with a reduced diffusion. This suggests that a
general temperature is appropriate in this aging regime. As the
system ages, the diffusion decreases further until $t_{w}
\approx$~100 min, when the diffusion of the 50 nm spheres no
longer decreases at the same rate as the 100 nm and 200 nm
spheres. At still later times, when $t_{w} \approx$~146 min, the
diffusion of the 100 nm spheres no longer decreases at the same
rate as the 200 nm spheres. This suggests that the Laponite
platelet structure begins to show inhomogeneities with age. The
smaller probes (which diffuse faster) are the first to sense these
inhomogeneities.

\begin{figure}
\epsfig{file=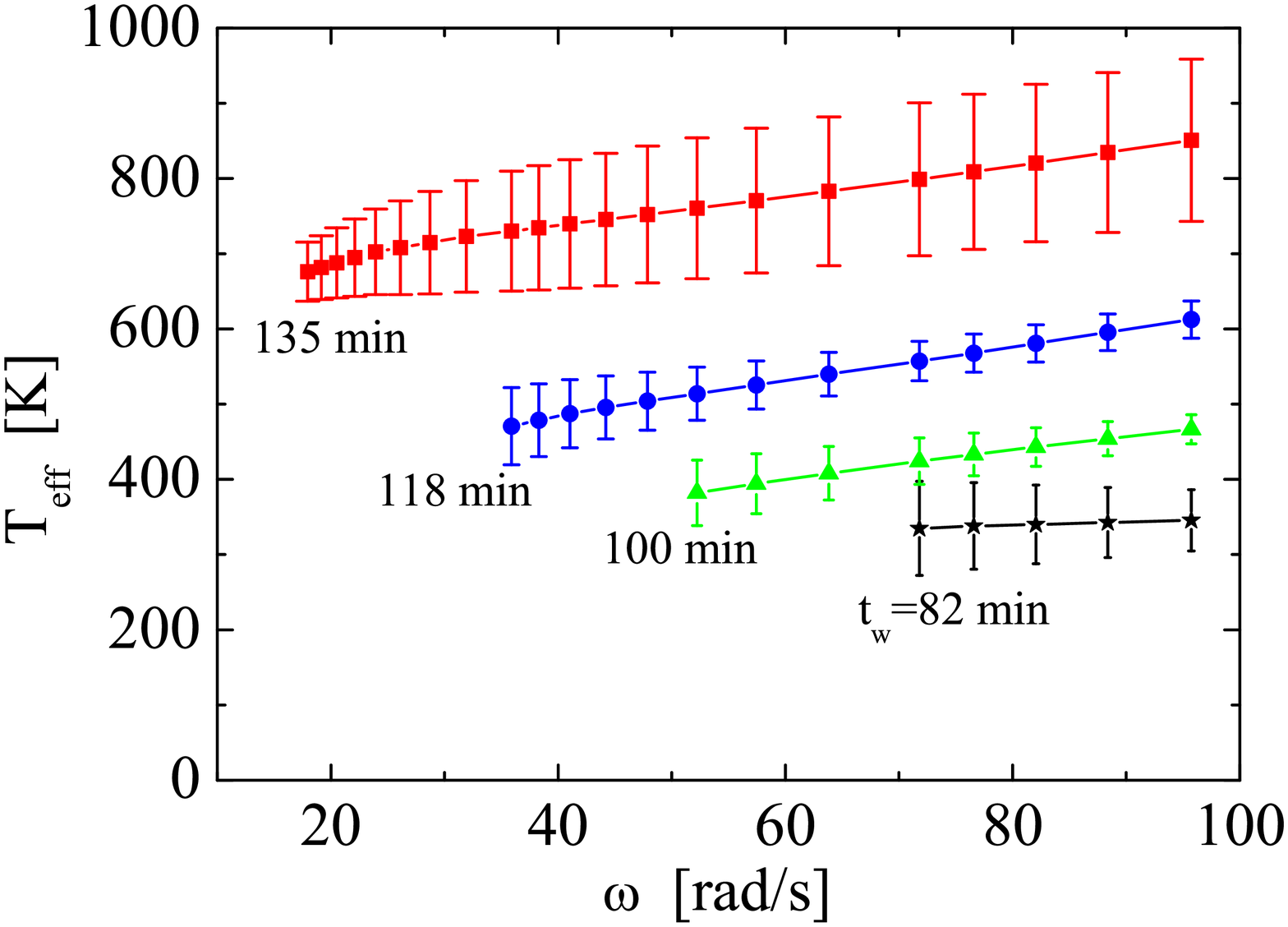,clip=,width=\columnwidthstrach}
\caption{\label{T_eff_fig} $T_{eff}$ determined from the
rheological measurements in Fig.\ \ref{visco-elastic} and the
time-dependent diffusion of 200 nm diameter probe particles shown
in Fig.\ \ref{scale_Dt}.}
\end{figure}

Having demonstrated agreement with an intensive general
temperature up to certain aging times, we now determine the
$T_{eff}$. To do this we numerically calculate the Laplace
transform of $D(t)$ and fit the term ${{k_B} / {6 \pi}
a\tilde{D}(s)}$ to a fourth order expansion of $\ln(s)$.
Substituting $i \omega$ for $s$ yields ${{k_B} / {6 \pi} a
D(\omega)}$. This, along with the measured $G^{*}(\omega)$ from
Fig.\ \ref{visco-elastic}, can be directly related to
$\Theta(\omega)$ through ${6 \pi} a D(\omega) G^{*}(\omega) /
{k_B} = \Theta(\omega)$, which in turn can be used to determine
$T_{eff}$ from Eq.\ (\ref{T_eff}). We limited the analysis to
$\mathrm{t_{w}} \leq 135$ min where $D(t)$ for the 100 nm and 200
nm spheres showed reasonable agreement, and to frequencies where
$D(\omega)$ could be determined for both sizes. Figure\
\ref{T_eff_fig} shows the frequency dependence of $T_{eff}$. To
determine this, we obtained $T_{eff}$ for the 100 and 200 nm
spheres with $\tilde{D}(s)$ determined with a Laplace transform
only over the experimental time scales. A third $T_{eff}$ was
determined from the 200 nm spheres by taking the transform up to
$t = \infty$, assuming $D(t)$ is constant for $t$ beyond the
experimental range. All three curves for $T_{eff}$ show the same
general trends, with the average of the three plotted in Fig.\
\ref{T_eff_fig} and their standard deviation represented by error
bars. For the aging times and frequencies accessible to our
experiment, there is a systematic rise of $T_{eff}$ in Fig.\
\ref{T_eff_fig} with aging time and with frequency, starting near
room temperature of 298 K. (For a $t_{w}$ of 82 min, $T_{eff}$
shows only a weak frequency dependence and its value is $335 \pm
63$ K.) This behavior is distinctly different to some other
determinations of effective temperatures in soft matter, which
have the opposite trend in aging time and
frequency~\cite{Bellon01_1el,Bonn03_1jcp}. Yet, our increase of
$T_{eff}$ with temperature seems to corroborate the short
aging-time behavior of Abou \emph{et al.}~\cite{Abou04_1prl}. In
that work, Abou \emph{et al.} found $T_{eff}$ to increase from
room temperature as a function of $t_{w}$ and then to decrease at
longer $t_{w}$. To account for this, they proposed that there were
three regimes: when the system is probed at a time scale (1) less
than, (2) equal to, and (3) greater than the characteristic
relaxation time of the Laponite system which increases with
$t_{w}$. For cases (1) and (3) Abou \emph{et al.} argue that the
system can thermalize to the bath and so $T_{eff}$ is
approximately room temperature, whereas for case (2) the Laponite
dispersion behaves as an out-of-equilibrium system with an
increased $T_{eff}$ when driven at its characteristic relaxation
frequency. Contrary to this view, our results suggest that there
are instead only two regimes -- for probed time scales (1) less
than and (2) greater than the characteristic relaxation time of
the Laponite. This stems from the fact that we find only a rise in
$T_{eff}$ without a subsequent decrease~\cite{aging_times}. In our
picture, the Laponite system is in thermal equilibrium with the
bath when probed at time scales greater than its characteristic
relaxation time. At probed time scales less than this relaxation
time, the system cannot thermalize and behaves as an
out-of-equilibrium system with a $T_{eff}$ greater than the bath
temperature.

In conclusion, we have measured the diffusion of probe particles
in an aging Laponite suspension which suggests that a general
temperature describes the system for short aging times. This is
indicated by the scaling of the diffusion with probe size, which
suggests that the system has a homogeneous viscoelastic behavior
on the length and time scales measured. Using our diffusion and
rheological measurements we determine a $T_{eff}(\omega)$ which
increases with frequency and aging time. This behavior suggests
two time-scale regimes; (1) for probing times greater than the
characteristic relaxation time of the glass where the system can
thermalize with the bath, and (2) for probing times less than the
characteristic relaxation time where the system is out of
equilibrium with the bath and has a $T_{eff} > T_{bath}$.

The authors acknowledge useful conversations with G. Bryant, M. A.
Anisimov, A. F. Kostko, and D. Garofalo.


\begin{thebibliography}{25}
\bibitem[{nan()}]{nano_spherres}
\bibinfo{note}{The 100 nm and 200 nm spheres were obtained from Duke Scientific
  suspended in an aqueous medium with 1 wt \% polystyrene spheres. The 50 nm
  spheres used to determine $T_{eff}$ were obtained as a 10 wt \% polystyrene
  sphere dispersion to reduce the amount of added aqueous solution to the
  Laponite.}
\bibitem[{con()}]{concentrations}
\bibinfo{note}{The concentrations are determined by taking the mass of added
  spheres divided by the mass of the Laponite dispersion, not including the
  aqueous solution accompanying the polystyrene spheres.}
\end{thebibliography}
\end{document}